\documentclass[aps,pre,twocolumn,showpacs,groupedaddress]{revtex4-1}

\usepackage[latin1]{inputenc}
\usepackage[T1]{fontenc}
\usepackage{graphicx}
\usepackage{amssymb}
\usepackage{amsmath}
\usepackage{wasysym}
\usepackage{xcolor}
\usepackage[bbgreekl]{mathbbol}
\usepackage{lettrine}
\usepackage{ulem} 
\usepackage{hyperref} 
\hypersetup{pdftitle={Article}, pdfauthor={Author}}

\def \et {\mathbf{e_{\theta}}}

\begin{document}
\title{Rheology of Granular Rafts}

\author{J. Lalieu}
\affiliation{Universit\'e Paris-Saclay, CNRS, FAST, 91405, Orsay, France}
\author{A. Seguin}
\affiliation{Universit\'e Paris-Saclay, CNRS, FAST, 91405, Orsay, France}
\author{G. Gauthier}
\affiliation{Universit\'e Paris-Saclay, CNRS, FAST, 91405, Orsay, France}

\date{\today}

\begin{abstract}
Rheology of macroscopic particle-laden interfaces, called ``{\it Granular Rafts}'' has been experimentally studied, in the simple shear configuration. The shear-stress relation obtained from a classical rheometer exhibits the same behavior as a Bingham fluid and the viscosity diverges with the surface fraction according to evolutions similar to 2D suspensions. The velocity field of the particles that constitute the granular raft has been measured in the stationary state. These measurements reveal non-local rheology similar to dry granular materials. Close to the walls of the rheometer cell, one can observe regions of large local shear rate while in the middle of the cell a quasistatic zone exists. This flowing region, characteristic of granular matter, is described in the framework of an extended kinetic theory showing the evolution of the velocity profile with the imposed shear stress. Measuring the probability density functions of the elementary strains, we provide evidence of a balance between positive and negative elementary strains. This behavior is the signature of a quasistatic region inside the granular raft.
\end{abstract}

\maketitle

\indent Particle-laden interfaces are ubiquitous in natural environment ({\it e.g.} insect colonies \cite{Loudet2005,Mlot2011}) and industries to build materials with specific properties ({\it e.g.} electric or magnetic \cite{Forth2019}), prevent sloshing \cite{Moisy2018} or stabilizing foams or emulsions \cite{Pickering1907,Binks2006}. Among their intriguing behavior one can cite their ability to generate armored non spherical or everlasting bubbles~\cite{Roux2022} which can support high over or under pressure \cite{Subramaniam2005,Timounay2017}. Their countless applications have generated many studies of their mechanical properties \cite{Pitois2019}.
When particles are spread on a liquid surface they deform the interface, interaction forces between them appear. For spheres, the contact angle $\zeta$ sets the position of attachment of the liquid/air interface on the particle. For light (small Bond number \cite{Vella2005}) and large enough particles, gravity and collo\"idal interactions are negligible compared to capillaries. The curvature of the interface results from the balance between the particle weight, the Archimedean force and the capillary force which pulls the beads, leading to attractive forces between particles. Under these conditions particle-laden interfaces are often called granular rafts \cite{Cicuta2009} and their ability to float, to sink or to trap material~\cite{Protiere2017, Lagarde2020} as well as their robustness has been widely studied \cite{Petit2016,Planchette2018}. The behavior of particle-laden interfaces under compression~\cite{Zang2010,Jambon2017,Saavedra2018}, or indentation~\cite{He2020} is fairly well understood. Their viscoelastic behavior has been studied \cite{Zang2010,Barman2014,Barman2016}, suggesting the importance of local interactions between particles in the macroscopic rheology, however their behavior when submitted to simple shear remains poorly understood.\\
In 3D, above a yield stress and in the inertial regime, dense granular materials (respectively suspensions) obey to the so-called local friction constitutive law $\mu(I) = \tau / P$ (resp. $\mu(J) = \tau / P$), where $\tau$ is the shear stress and $P$ the confining pressure, and a dilatancy law $\phi(I)$ (resp. $\phi(J)$) where $\phi$ is the packing fraction. $\mu$ and $\phi$ are scalar functions of the inertial number $I=\dot{\gamma}_{\ell} d / \sqrt{P /\rho}$ (resp. the viscous number $J=\eta_f \dot{\gamma}_{\ell} / P $), with $\dot{\gamma}_{\ell}$ the local shear rate, $d$ the particle diameter, $\rho$ their density (and $\eta_f$ the fluid viscosity) \cite{GDR2004,Cassar2005,Jop2006,Boyer2011}. This inertial (resp. viscous) number can be seen as the ratio between a characteristic time of strain $1/\dot{\gamma}_{\ell}$ and a characteristic time of rearrangement $d\sqrt{\rho / P}$ (resp. $\eta_f / P$). However in many cases granular materials exhibit non-local effects which lead to the development of a sheared region next to a quasistatic one. In these situations, granular material rheology deviates from the local constitutive law and several models have been developed to account for this nonlocality \cite{Bocquet2009,Reddy2011,Bouzid2013,Bouzid2015,Henann2014,Thomas2019,Gaume2020}.\\ 
\indent To keep the same rheological framework for the description of rafts, we define a microscopic characteristic time $t_c$ related to the rearrangement of the constituting grains. To define $t_c$, the useful stress scale comes from the surface tension $\chi$ between the liquid and the particles which apply the confining pressure $\sigma/d$ with $\sigma=\chi \cos(\zeta)$. Thus one might define this time as $t_c = d \sqrt{ \rho d / \sigma }$ and build a {\it capillary inertial number}: $I_c = \dot{\gamma}_{\ell}d/ \sqrt{ \sigma /  \rho d}$.
Moreover, due to the capillary forces between the particles, granular rafts may belong to the attractive granular class of materials \cite{Lois2008} with stronger non-local effects, thus different rheology.\\
\indent In this letter we study the rheology of a granular raft with a classical rheometer with imposed shear stress and address the question of the locality of the particle-laden interfaces behavior. Granular rafts are two-dimensional attractive granular media that exhibit a yield stress function of the particle surface fraction and whose mean behavior can be described as a Bingham fluid. When the raft is sheared, coupling global stress-strain measurements and displacement field measurements reveals that the local capillary inertial number $I_c$ is not homogeneous. While accounting for this behavior using the framework of a continuous hydrodynamic model based on the kinetic theory extended to dense granular systems \cite{Losert2000, Gaume2020}, we highlight that in the region where $I_c$ is smaller than a critical inertial number ${I_c}^{\ast}$, the microscopic velocity fluctuations gives rise to a balance between positive and negative elementary strains, thus inhibiting the onset of a macroscopic shear. This is characteristic of a quasistatic regime and suggests a transition to elasticity in nature of the contacts between grains. \\
\begin{figure}[t!]
\begin{center}
\includegraphics[width=\linewidth]{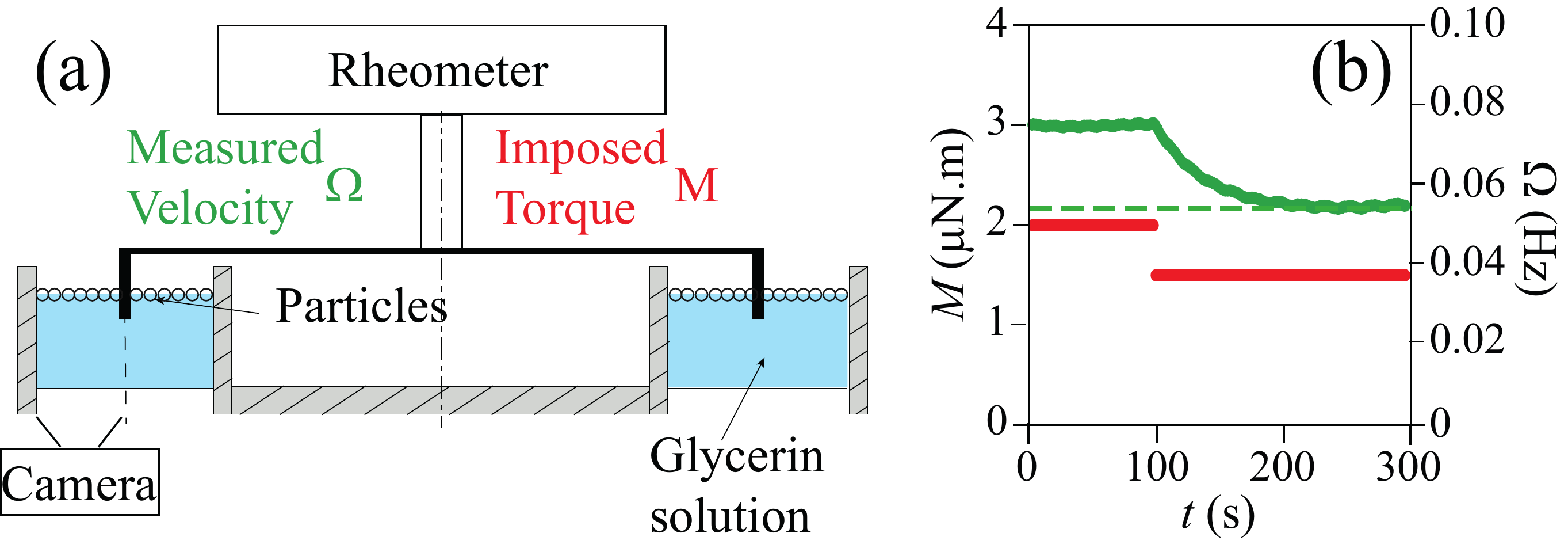}
   \caption{(a) Sketch of the experimental setup.
   (b) Imposed torque $M=1.5$~$\mu$N.m ({\color{red}{$\bullet$}}) and measured velocity of the cylinder $\Omega$ ({\color{green}{$\bullet$}}) as a function of time $t$ for  $\phi = 0.74$. The dotted line represents the steady regime with $\Omega_{\infty} = 54$~mHz.}
\label{setup}
\end{center}
\end{figure}

\begin{figure}[t!]
\begin{center}
\includegraphics[width=\linewidth]{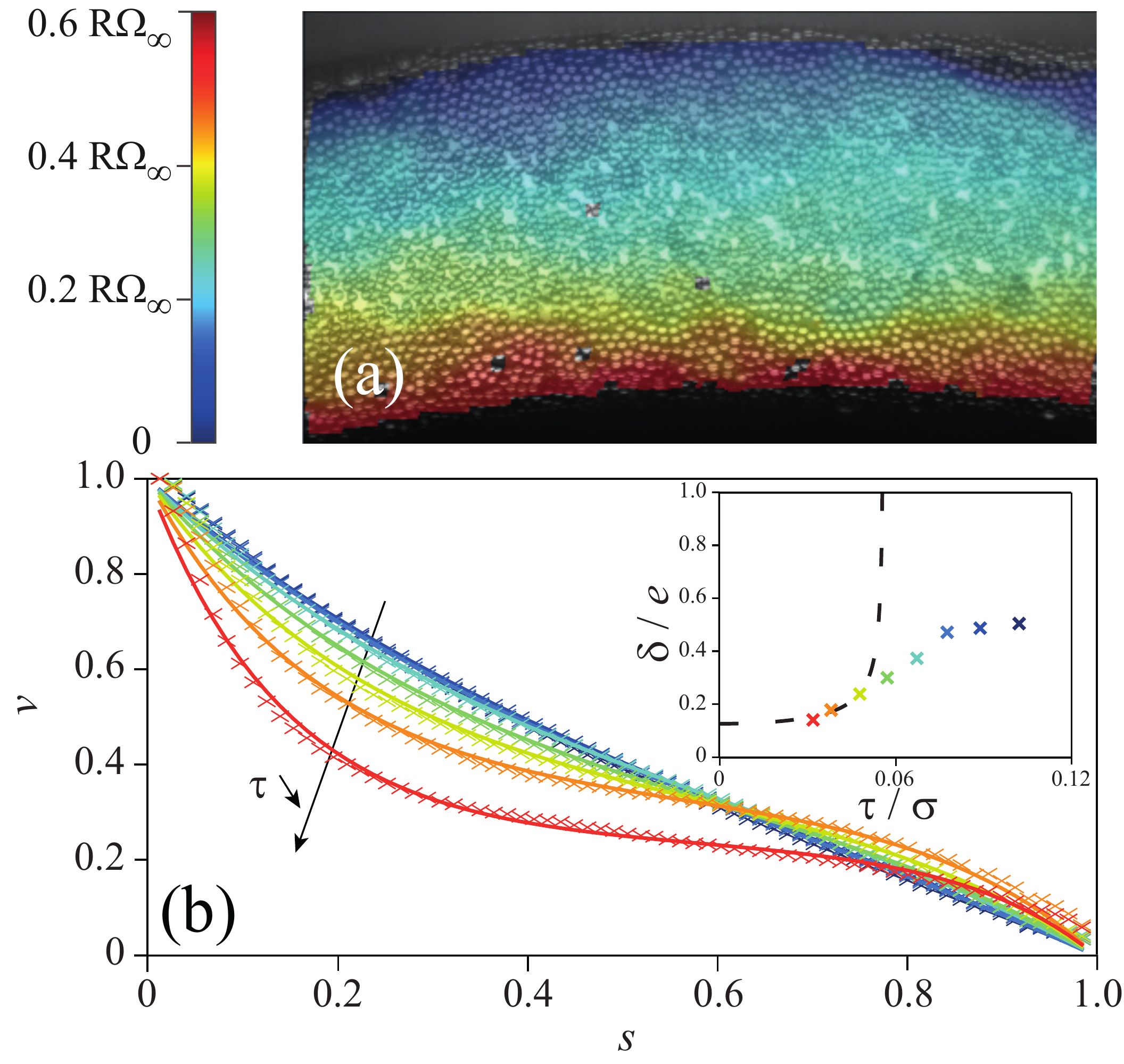}
   \caption{(a) Typical instantaneous velocity field for an imposed torque $M=6$~$\mu$N.m. The color of the velocity vector represents its norm relatively to the velocity of the inner cylinder ($R\Omega_{\infty}=17.5 $~mm.s$^{-1}$).
   (b) Normalized velocity profile $v$ as a function of the distance $s$ inside the gap. Same symbols as in figure \ref{rawdata}. The solid lines are given by Eq.~\ref{analv2}. Inset: characteristic length $\delta/e$ as a function of $\tau /\sigma$. The dotted line is obtained from the hydrodynamical model (Eq.~\ref{analv}) with $(2\kappa_0\eta_0)^{1/2}=12.6 \times 10^{-2}$ Pa.m$^2$ and $(\epsilon_0\eta_0)^{1/2}=3.3\times 10^{-4}$ Pa.m.} 
\label{veloc}
\end{center}
\end{figure}

\indent \textit{Experimental Setup} -- Granular rafts are obtained by spreading silanized polystyrene spheres (diameter $d=140$~$\mu$m, wetting contact angle $\zeta \approx 80^\circ$), setting the mean surface fraction $\phi$, over a mixture of water and glycerin matching particle density as to avoid sedimentation. Tetradecyl trimethyl ammonium bromide (TTAB) is added to the liquid phase (concentration  $10$~g.L$^{-1}$) to reduce the cohesive force between the particles, leading to $\sigma \approx 5.9$~mN.m$^{-1}$.\\
\indent Rafts are sheared in a cylindrical double gap homemade cell of mean radius $R=30$~mm whose two gaps are $e = 4.5$~mm wide (Fig~\ref{setup}a). All the walls are made coarse by gluing at their surfaces the same particles as the ones forming the raft. We place the cell into a MCR 501 rheometer and lower the measuring cylinder $10$~mm deep into the solution. The raft is sheared at constant velocity for ten rotations before any measurement. The cylinder is then driven with a constant torque $M$ and we allow the system to flow until a steady state is achieved, with a constant rotational speed $\Omega_{\infty}$ measured, which comes typically in $150$~s (Fig.~ \ref{setup}b) and is driven by the fluid flow underneath the raft. Note that the rotational speed $\Omega$ corresponds to the cylinder velocity. This steady state can be achieved both with decreasing or increasing torque, showing no hysteresis or long-time variation, and a benchmark measure is performed at the same depth without particles to obtain the resisting torque $M_f$ for the pure fluid at $\Omega_{\infty}$. The expression of the shear stress on the raft, that is the shear stress integrated over its thickness, is then $\displaystyle \tau = (M-M_f)/(4\pi R^2)$.\\
\indent Using a camera set under the raft, the displacement field in the outer gap of the cell is recorded while shearing (Fig.~\ref{veloc}a), then processed through image correlation to compute the local time-averaged velocity field. Considering the flow geometry, the results will be presented in polar coordinates $(r,\theta)$ centered on the axis of the rheometer in the viewing plane of the camera. Given the range of variation of $r$, we define a reduced space variable $s=(r-R) / e$. Additionally, the axisymmetry of the system allows to average along the orthoradial direction $\et$. Moreover, once in the steady state, we do not observe any displacement along the radial direction. Thus, the instantaneous velocity is $V_{\theta}(s,t)\et$. Averaging over time, one obtains the time-averaged velocity $v_{\theta}(s)=<V_{\theta}(s,t)>$. It is then possible to determine the local shear rate $\dot{\gamma}_{\ell}(s)$. The detection of the grains is feasible in the center of the cell leading to no significant radial variations of $\phi$ in the range $0.3<s<0.9$, implying no variation in the whole gap. \\

\begin{figure}[t!]
\begin{center}
\includegraphics[width=\linewidth]{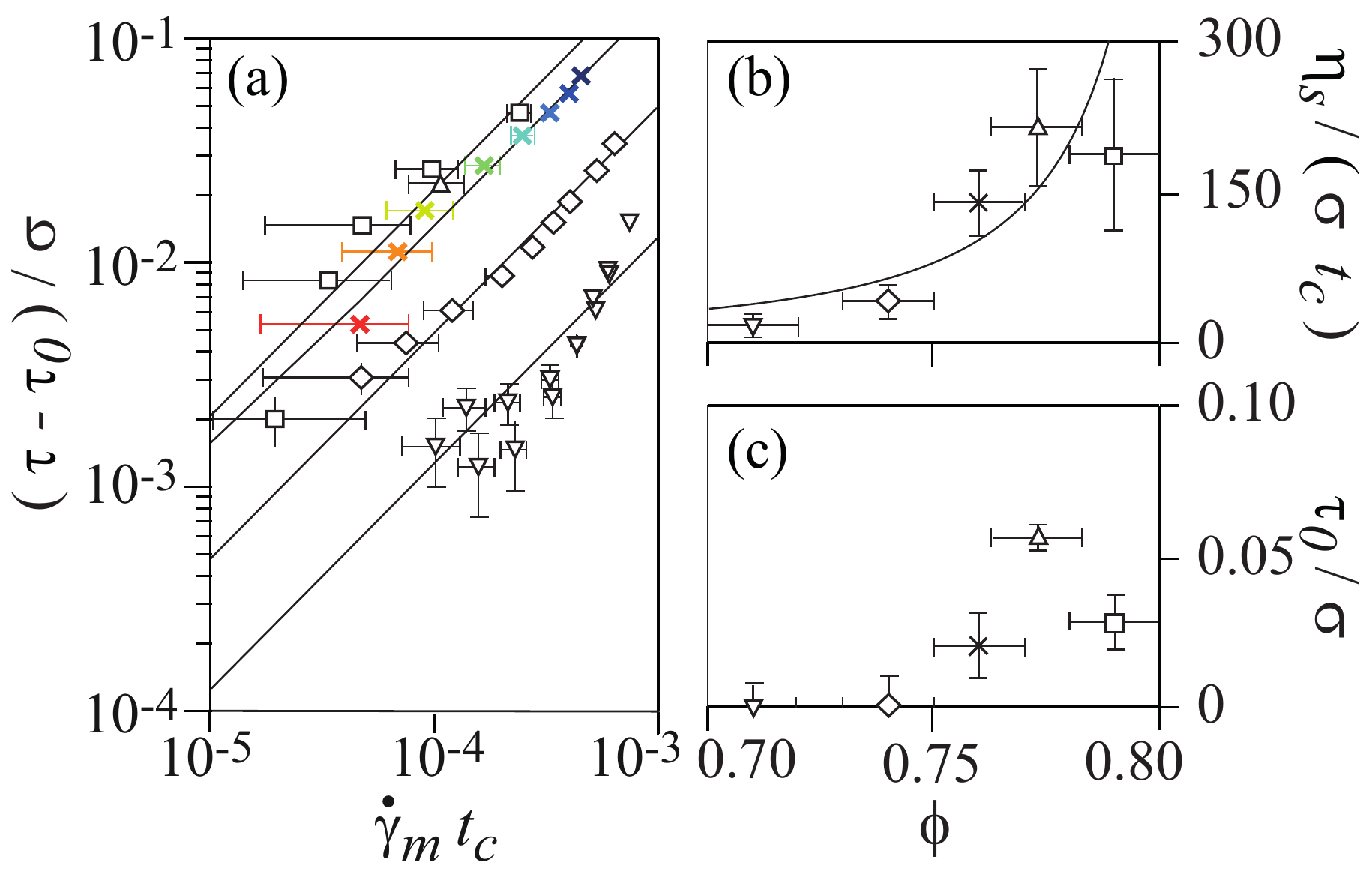}
   \caption{ (a)  Dimensionless shear stress $(\tau-\tau_0)/\sigma$ as a function of the dimensionless mean strain $\dot{\gamma}_m t_c$ for different solid fractions: ($\triangledown$) $\phi=0.71$, ($\diamond$) $\phi=0.74$, ($\times$) $\phi=0.76$, ($\square$) $\phi=0.79$ and ($\triangle$) $\phi=0.77$. The solid line represents a linear fit $\tau-\tau_0=\eta_s(\phi)\,\sigma t_c \dot{\gamma}_m$ corresponding to Bingham fluid behavior.
   (b) Normalized surface viscosity $\eta_s / (\sigma\, t_c ) $ and (c) dimensionless yield stress $\tau_0/\sigma$ as a function of $\phi$. The solid line in (b) is given by $\eta_s/ (\sigma\, t_c )  = \eta_0 \phi_c \left(\phi_c-\phi\right)^{-2\phi_c}$ with $\phi_c=0.82$ and $\eta_0  = 1.22$ corresponding to 2D suspension behavior law \cite{Krieger1972, Brady1983}. }
   \label{rawdata}
   \end{center}
\end{figure}


\indent \textit{Rheometry} -- In these experiments performed in a Couette rheometer, it is usual to present the evolution of the mean (surface) stress $\tau$ as a function of the mean shear rate $\dot{\gamma}_m=R\Omega_{\infty}/e$ (Fig.~\ref{rawdata}). The averaged rheological curves show that the rheology of the granular raft follows a Bingham fluid constitutive law $\displaystyle \tau = \tau_0 + \eta_s \dot{\gamma}_m$ \cite{Bingham1922}, where $\tau_0$ is a 2D surface yield stress in Pa.m and $\eta_s$ is then a 2D surface viscosity thus expressed in Pa.m.s. Figure \ref{rawdata}a presents the linear evolution of the normalized stress $(\tau - \tau_0)/\sigma$ as a function of the normalized shear rate $\dot{\gamma}_m t_c$ for different packing fractions $\phi$. Above a critical particle surface fraction $\phi^{\ast} \approx 0.71$ granular rafts exhibit a yield stress $\tau_0$ which increases with $\phi$ (Fig.~\ref{rawdata}c). Flowing rafts do so with a constant viscosity, which is a growing function of $\phi$ (Fig.~\ref{rawdata}b). These findings are in a roughly good agreement with previous studies \cite{Reynaert2007, Zang2010} and follow the usual rheological law of 2D suspensions $\eta_s \propto \left(\phi_c-\phi \right)^{-2\phi_c}$ \cite{Krieger1972,Brady1983}.


\indent \textit{Velocity field} -- The images taken from a video camera are analyzed by a DIC software (DaVis, LaVision) to get the velocity field of the grains (Fig~\ref{veloc}a). From it, we extract azimutal profile $v_{\theta}$. In the observed range, the velocity at which the cylinder rotates $R\Omega$ is never met by the grains at the wall. To account for this slip velocity, we normalize the velocity by its maximum value $v_{M}$ leading to $v=v_{\theta}/v_{M}$. Figure \ref{veloc}b shows the local velocity $v$ measurements as a function of distance $s$ from the cylinder for decreasing imposed shear stress $\tau$. Overall, we see that the velocity decreases as the distance to the inner cylinder increases. From this velocity field, we can deduce the local shear rate $\dot{\gamma}_\ell=r~d ( v_\theta /r) / d r \simeq d v_\theta / d r $ in our experiments. This decrease of $v$ is rather linear when the applied stress $\tau$ is high, leading to a roughly constant shear rate $\dot{\gamma}_\ell$ in the raft. But it becomes nonlinear as $\tau$ becomes smaller. We observe a localization of the velocity close to the wall like what can be sometimes observed in dry granular media \cite{Losert2000,Bocquet2001,Reddy2011,Seguin2011}. Thus, the local shear rate $\dot{\gamma}_\ell$ is not homogeneous in the raft and these velocity field measurements show that the rheology of the rafts is expected to be non-local, different from a Newtonian fluid \cite{Guyon2001}, a suspension \cite{Blanc2011} and dry granular medium \cite{Reddy2011}.\\
\begin{center}
\begin{figure*}[t!]
\includegraphics[width=\linewidth]{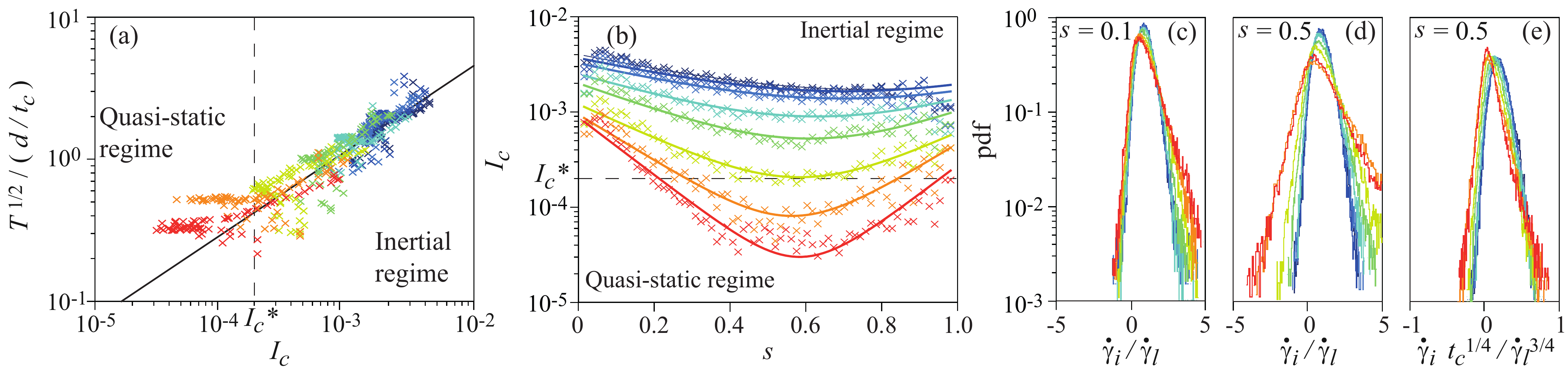}
   \caption{(a) Dimensionless velocity fluctuations $T^{1/2}/(d/t_c)$ as a function of $I_c$ for different imposed $\tau$ and for $\phi=0.76$ (same symbols as in Fig~\ref{rawdata}). The solid line represents the best fit of the data $T^{1/2} \sim I_c^{1/(2\beta-1)}$ with $\beta=1.25\pm0.05$.
   (b) $I_c$ as a function of $s$. The dotted line represents the critical value $I_c^{\ast}$ delimiting the two flow regimes.
    PDF of elementary strains $\dot{\gamma}_i$ normalized: by the mean local strain $\dot{\gamma}_{\ell}$ in the flowing region (c) and quasistatic region (d); by $\dot{\gamma}_{\ell}^{3/4}$ in the quasistatic region (e).}
   \label{Temp}
\end{figure*}
\end{center}
\indent \textit{Hydrodynamic Model} -- To account for non-locality, the recent rheological models applied to granular flows define a diffusive quantity. Even though the most universal one (in its application) is the non-local granular fluidity (defined as $\dot{\gamma}_\ell/\mu$), a recent review \cite{Kamrin2019} suggests that kinetic theory can be successfully applied while also giving a microscopic physical origin for the velocity fluctuations. Thus, the kinetic theory model is both relevant and sufficient in the case of an homogeneous state of stress. \\
\indent We develop a hydrodynamic model based on the kinetic theory for dry granular media \cite{Losert2000,Bocquet2001,Seguin2011,Seguin2017}. The classical kinetic theory of molecular systems has been applied with some success to dilute and even dense athermal granular systems by introducing the concept of a "temperature" $T$ related to the fluctuations of the time-averaged velocity $T (s)= <V_\theta(s,t)^2>-v_\theta(s)^2$. Within this framework, heat is created by the flow itself. A moving area increases locally the temperature, thus reducing the resistance to movement of the surrounding particles and allowing them to flow. This effect is then propagated until a steady state is reached. In the present case, the local velocity fluctuations are generated and exchanged in the whole raft through the contacts of the particles in the flow. Assuming pressure $p$ and $\tau$ are homogeneous in the whole raft, the effective viscosity varies such as $\eta \propto 1/\dot{\gamma}_\ell$. In the framework of the kinetic theory of granular systems \cite{Bocquet2001}, we can define an effective surface viscosity $\eta $ related to the temperature $T$ such as $\eta = \eta_0 T^{-(2\beta-1)/2}$ where $\eta_0$ depends on density, diameter, mechanical properties of the particles and pressure, which are constant in the experiment, and $\beta$ is a phenomenological exponent equal to one in dilute and moderately dense systems, and bigger than one in highly dense systems to account for the divergence of the viscosity, e.g. $\beta \simeq 1.75$ for granular shear flow \cite{Losert2000,Bocquet2001}. The relation between $\eta$ and $T$ implies a power-law between the temperature and the local shear-rate $\dot{\gamma}_{\ell}$. It holds in the inertial regime, that is as long as the contacts by collisions are dominant in the dynamics of the raft; it is then useful to plot the local temperature $T$ as a function of the local capillary inertial number $I_c$ (Fig.~\ref{Temp}a). For high capillary inertial number, we observe a quasi-linear relationship between the two quantities, with $\beta \simeq 1.25$. To latter obtain an analytical solution while introducing no noticeable error, we set $\beta=1$ leading to $I_c = \dot{\gamma}_\ell t_c = \tau T^{1/2} t_c /\eta_0$, which shows this framework is compatible with the measurements in this flowing regime. For low capillary inertial number, there is another regime where this relationship is no longer valid. It is then possible to define a crossover between these two, characterized by a critical inertial capillary number $I_c^\ast = 2 \times 10^{-4}$. Measures of $I_c$ inside the raft (Fig.~\ref{Temp}b) show that this criterion is met everywhere for the higher stresses, whereas for lower applied stresses $I_c$ falls under the criterion at a position $s=s^{\ast}$, which we read from figure \ref{Temp}b and define as the position at which $I_c(s^{\ast})=I_c^{\ast}$.\\
\indent Above $I_c^{\ast}$, the temperature obeys the heat equation in our  configuration: equilibrium between diffusion (with a transport coefficient reducing to $\kappa = \kappa_0 T^{-1/2}$), collision dissipation (with a coefficient reducing to $\varepsilon = \varepsilon_0 T^{-1/2}$) and source term corresponding to $\tau \dot{\gamma}_{\ell}$. $\kappa_0$ and $\varepsilon_0$ depend on density, diameter and mechanical properties of the grains and pressure $p$ which are constant in our case. In our geometry, the hydrodynamic equation for $T(s)$ comes down to:
\begin{equation}
\frac{d}{ds}\left(\kappa(T)\frac{dT}{ds} \right)- \varepsilon(T) T + \frac{\tau^2}{\eta_0} T^{1/2}=0.
\label{heat}
\end{equation}
which can be integrated to obtain: 
\begin{equation}
\frac{d^3 v}{ds^3} - \frac{e^2}{\delta^2} \frac{d v}{ds} =0.
\label{analv}
\end{equation}
where $\delta = ((2\kappa_0 \eta_0) /(\varepsilon_0 \eta_0 -\tau^2))^{1/2}$ is a characteristic length. Solving equation \ref{analv} using $v(0)=1$ and $v(1)=0$ gives:
\begin{widetext}
\begin{equation}
 \displaystyle v(s)=  A \left(\cosh\left(\frac{(2s-1)e}{2\delta}\right)-\cosh\left(\frac{e}{2\delta}\right)\right) \\
   +  \exp\left(- \frac{se}{2\delta}\right) \frac{ \displaystyle \sinh\left(\frac{(1-s)e}{2\delta}\right)}  {\displaystyle \sinh\left(\frac{e}{2\delta}\right)} 
    \label{analv2}
\end{equation}
\end{widetext}

\noindent with A a fitting parameter. This analytical function has been fitted on the velocity profiles showing a excellent agreement with the experimental data (Fig.~\ref{setup}d). Note that the velocity profile so obtained presumes that the whole raft is in an inertial regime. These fits are robust and can be derived in order to extend them to the $I_c$ profiles (Fig.~\ref{Temp}b). The diffusion length $\delta$, found in the analytical solution for $v$, is a growing function of $\tau$, as observed in figure \ref{Temp}c. Its evolution differs however from the analytical model possibly through the finite size of the experimental system ($\delta /e \sim 1 $). Consequently, the stress exceeds the maximal value for an infinite system and estimated at $(\epsilon_0\eta_0)^{1/2}/\sigma=0.056$.\\ 

\indent \textit{Quasistatic Regime} -- Despite the good agreement between velocity profiles deduced from equation~\ref{analv2} and the experimental data, below the critical inertial capillary number $I_c^{\ast}$ the system is no longer described with the hydrodynamical model. According to a recent numerical study~\cite{Gaume2020}, for $I_c(s)<I_c^\ast$, the system is in a quasistatic regime (similar to a plug flow) in which the strain and thus the velocity fluctuations are sustained by the boundary conditions. In this regime, elementary (local and instantaneous) strains occur over time in and against the forcing. The Probability Density Function of the elementary strains $\dot{\gamma}_i$ normalized with the local shear rate $\dot{\gamma}_l$ for two radial locations $s=0.1$ and $s=0.5$ are displayed in figures \ref{Temp}c and \ref{Temp}d. For $s=0.1$, $I_c > I_c^{\ast}$ for any $\tau > \tau_0$, PDF are narrow and the elementary strains are positive, \textit{i.e.} in the direction of $\dot{\gamma}_{\ell}$ (Fig.~\ref{veloc}b). This is in agreement with a predominance of a viscous component of the stress over an elastic one, as developed in an other numerical study \cite{Tighe2010}. At $s=0.5$, for the higher imposed torques ($\tau/\sigma \geq 5.7 \times 10^{-2}$), $I_c > I_c^{\ast}$ and the PDF are similar. It is no more the case at lower imposed torques, for which $I_c \leq I_c^{\ast}$: the PDF are large and present negative values (fig~\ref{Temp}d); showing important elementary strains opposed to the shear flow, characteristic of the quasistatic regime \cite{Gaume2020}. Furthermore, while the velocity profiles cannot show it, the PDF show that in the quasistatic regime, the particles interact through elastic contacts. Indeed, once normalized by  ${\dot{\gamma}_l}^{3/4}$, the PDF collapse on a single curve (fig~\ref{Temp}e), as proposed in \cite{Tighe2010}. This behavior is a signature of the quasistatic regime. \\


\indent \textit{Concluding remarks} -- The mechanical behavior of granular rafts is close to the one of attractive granular materials. The flow exhibits a non uniform shear, thus revealing two regimes separated by a critical value of the local capillary inertial number $I_c$. A hydrodynamical model developed from kinetic theory describes well the flow and as predicted by a recent numerical model \cite{Gaume2020}, below the critical value $I_c^\ast$ the system is in a quasistatic regime. The signature of this quasistatic regime is reflected in the scalings of the probability density function of the elementary strain $\dot{\gamma_i}$, as proposed by an other numerical model \cite{Tighe2010}. This suggests that the macroscopic transition in the flow regime is in agreement with a microscopic transition in the nature of contacts between the particles. The critical value $I_c^\ast \approx 2 \times 10^{-4}$ is found to be one order of magnitude lower than the one predicted for 2D cohesion-free granular materials ($I^\ast \approx 5\times 10^{-3}$), in agreement with numerical study which reported that characteristic relaxation time $t_c$ can be two orders of magnitude lower for attractive granular materials \cite{Chaudhuri2012}.\\

\begin{acknowledgments}
We are grateful to L. Auffray, J. Amarni, A. Aubertin, C. Manquest and R. Pidoux for the development of the experimental setup. The authors thanks N. Retailleau, Y. Khidas, O. Pitois and F. Rouyer for fruitfull discussions. This work is supported by the project PhyGaMa ANR-19-CE30-0009-02 and "Investissements d'Avenir" LabEx Physique:
Atomes Lumi\`ere Mati\`ere (ANR-10-LABX-0039-PALM).
\end{acknowledgments}

\bibliography{main-arXivRaft}

\begin{thebibliography}{47}%
\makeatletter
\providecommand \@ifxundefined [1]{%
 \@ifx{#1\undefined}
}%
\providecommand \@ifnum [1]{%
 \ifnum #1\expandafter \@firstoftwo
 \else \expandafter \@secondoftwo
 \fi
}%
\providecommand \@ifx [1]{%
 \ifx #1\expandafter \@firstoftwo
 \else \expandafter \@secondoftwo
 \fi
}%
\providecommand \natexlab [1]{#1}%
\providecommand \enquote  [1]{``#1''}%
\providecommand \bibnamefont  [1]{#1}%
\providecommand \bibfnamefont [1]{#1}%
\providecommand \citenamefont [1]{#1}%
\providecommand \href@noop [0]{\@secondoftwo}%
\providecommand \href [0]{\begingroup \@sanitize@url \@href}%
\providecommand \@href[1]{\@@startlink{#1}\@@href}%
\providecommand \@@href[1]{\endgroup#1\@@endlink}%
\providecommand \@sanitize@url [0]{\catcode `\\12\catcode `\$12\catcode
  `\&12\catcode `\#12\catcode `\^12\catcode `\_12\catcode `\%12\relax}%
\providecommand \@@startlink[1]{}%
\providecommand \@@endlink[0]{}%
\providecommand \url  [0]{\begingroup\@sanitize@url \@url }%
\providecommand \@url [1]{\endgroup\@href {#1}{\urlprefix }}%
\providecommand \urlprefix  [0]{URL }%
\providecommand \Eprint [0]{\href }%
\providecommand \doibase [0]{http://dx.doi.org/}%
\providecommand \selectlanguage [0]{\@gobble}%
\providecommand \bibinfo  [0]{\@secondoftwo}%
\providecommand \bibfield  [0]{\@secondoftwo}%
\providecommand \translation [1]{[#1]}%
\providecommand \BibitemOpen [0]{}%
\providecommand \bibitemStop [0]{}%
\providecommand \bibitemNoStop [0]{.\EOS\space}%
\providecommand \EOS [0]{\spacefactor3000\relax}%
\providecommand \BibitemShut  [1]{\csname bibitem#1\endcsname}%
\let\auto@bib@innerbib\@empty
\bibitem [{\citenamefont {Loudet}\ \emph {et~al.}(2005)\citenamefont {Loudet},
  \citenamefont {Alsayed}, \citenamefont {Zhang},\ and\ \citenamefont
  {Yodh}}]{Loudet2005}%
  \BibitemOpen
  \bibfield  {author} {\bibinfo {author} {\bibfnamefont {J.~C.}\ \bibnamefont
  {Loudet}}, \bibinfo {author} {\bibfnamefont {A.~M.}\ \bibnamefont {Alsayed}},
  \bibinfo {author} {\bibfnamefont {J.}~\bibnamefont {Zhang}}, \ and\ \bibinfo
  {author} {\bibfnamefont {A.~G.}\ \bibnamefont {Yodh}},\ }\href {\doibase
  10.1103/PhysRevLett.94.018301} {\bibfield  {journal} {\bibinfo  {journal}
  {Phys. Rev. Lett.}\ }\textbf {\bibinfo {volume} {94}},\ \bibinfo {pages}
  {018301} (\bibinfo {year} {2005})}\BibitemShut {NoStop}%
\bibitem [{\citenamefont {Mlot}\ \emph {et~al.}(2011)\citenamefont {Mlot},
  \citenamefont {Tovey},\ and\ \citenamefont {Hu}}]{Mlot2011}%
  \BibitemOpen
  \bibfield  {author} {\bibinfo {author} {\bibfnamefont {N.~J.}\ \bibnamefont
  {Mlot}}, \bibinfo {author} {\bibfnamefont {C.~A.}\ \bibnamefont {Tovey}}, \
  and\ \bibinfo {author} {\bibfnamefont {D.~L.}\ \bibnamefont {Hu}},\ }\href
  {\doibase 10.1073/pnas.1016658108} {\bibfield  {journal} {\bibinfo  {journal}
  {Proceedings of the National Academy of Sciences}\ }\textbf {\bibinfo
  {volume} {108}},\ \bibinfo {pages} {7669} (\bibinfo {year}
  {2011})}\BibitemShut {NoStop}%
\bibitem [{\citenamefont {Forth}\ \emph {et~al.}(2019)\citenamefont {Forth},
  \citenamefont {Kim}, \citenamefont {Xie}, \citenamefont {Liu}, \citenamefont
  {Helms},\ and\ \citenamefont {Russell}}]{Forth2019}%
  \BibitemOpen
  \bibfield  {author} {\bibinfo {author} {\bibfnamefont {J.}~\bibnamefont
  {Forth}}, \bibinfo {author} {\bibfnamefont {P.~Y.}\ \bibnamefont {Kim}},
  \bibinfo {author} {\bibfnamefont {G.}~\bibnamefont {Xie}}, \bibinfo {author}
  {\bibfnamefont {X.}~\bibnamefont {Liu}}, \bibinfo {author} {\bibfnamefont
  {B.~A.}\ \bibnamefont {Helms}}, \ and\ \bibinfo {author} {\bibfnamefont
  {T.~P.}\ \bibnamefont {Russell}},\ }\href {\doibase
  https://doi.org/10.1002/adma.201806370} {\bibfield  {journal} {\bibinfo
  {journal} {Advanced Materials}\ }\textbf {\bibinfo {volume} {31}},\ \bibinfo
  {pages} {1806370} (\bibinfo {year} {2019})}\BibitemShut {NoStop}%
\bibitem [{\citenamefont {Moisy}\ \emph {et~al.}(2018)\citenamefont {Moisy},
  \citenamefont {Bouvard},\ and\ \citenamefont {Herreman}}]{Moisy2018}%
  \BibitemOpen
  \bibfield  {author} {\bibinfo {author} {\bibfnamefont {F.}~\bibnamefont
  {Moisy}}, \bibinfo {author} {\bibfnamefont {J.}~\bibnamefont {Bouvard}}, \
  and\ \bibinfo {author} {\bibfnamefont {W.}~\bibnamefont {Herreman}},\ }\href
  {\doibase 10.1209/0295-5075/122/34002} {\bibfield  {journal} {\bibinfo
  {journal} {{EPL} (Europhysics Letters)}\ }\textbf {\bibinfo {volume} {122}},\
  \bibinfo {pages} {34002} (\bibinfo {year} {2018})}\BibitemShut {NoStop}%
\bibitem [{\citenamefont {Pickering}(1907)}]{Pickering1907}%
  \BibitemOpen
  \bibfield  {author} {\bibinfo {author} {\bibfnamefont {S.~U.}\ \bibnamefont
  {Pickering}},\ }\href@noop {} {\bibfield  {journal} {\bibinfo  {journal} {J.
  Chem. Soc., Trans}\ }\textbf {\bibinfo {volume} {91}},\ \bibinfo {pages}
  {2001} (\bibinfo {year} {1907})}\BibitemShut {NoStop}%
\bibitem [{\citenamefont {Binks}\ and\ \citenamefont
  {Murakami}(2006)}]{Binks2006}%
  \BibitemOpen
  \bibfield  {author} {\bibinfo {author} {\bibfnamefont {B.~P.}\ \bibnamefont
  {Binks}}\ and\ \bibinfo {author} {\bibfnamefont {R.}~\bibnamefont
  {Murakami}},\ }\href@noop {} {\bibfield  {journal} {\bibinfo  {journal}
  {Nature materials}\ }\textbf {\bibinfo {volume} {5}},\ \bibinfo {pages} {865}
  (\bibinfo {year} {2006})}\BibitemShut {NoStop}%
\bibitem [{\citenamefont {Roux}\ \emph {et~al.}(2022)\citenamefont {Roux},
  \citenamefont {Duchesne},\ and\ \citenamefont {Baudoin}}]{Roux2022}%
  \BibitemOpen
  \bibfield  {author} {\bibinfo {author} {\bibfnamefont {A.}~\bibnamefont
  {Roux}}, \bibinfo {author} {\bibfnamefont {A.}~\bibnamefont {Duchesne}}, \
  and\ \bibinfo {author} {\bibfnamefont {M.}~\bibnamefont {Baudoin}},\ }\href
  {\doibase 10.1103/PhysRevFluids.7.L011601} {\bibfield  {journal} {\bibinfo
  {journal} {Phys. Rev. Fluids}\ }\textbf {\bibinfo {volume} {7}},\ \bibinfo
  {pages} {L011601} (\bibinfo {year} {2022})}\BibitemShut {NoStop}%
\bibitem [{\citenamefont {Bala~Subramaniam}\ \emph {et~al.}(2005)\citenamefont
  {Bala~Subramaniam}, \citenamefont {Abkarian}, \citenamefont {Mahadevan},\
  and\ \citenamefont {Stone}}]{Subramaniam2005}%
  \BibitemOpen
  \bibfield  {author} {\bibinfo {author} {\bibfnamefont {A.}~\bibnamefont
  {Bala~Subramaniam}}, \bibinfo {author} {\bibfnamefont {M.}~\bibnamefont
  {Abkarian}}, \bibinfo {author} {\bibfnamefont {L.}~\bibnamefont {Mahadevan}},
  \ and\ \bibinfo {author} {\bibfnamefont {H.~A.}\ \bibnamefont {Stone}},\
  }\href@noop {} {\bibfield  {journal} {\bibinfo  {journal} {Nature}\ }\textbf
  {\bibinfo {volume} {438}},\ \bibinfo {pages} {930} (\bibinfo {year}
  {2005})}\BibitemShut {NoStop}%
\bibitem [{\citenamefont {Timounay}\ \emph {et~al.}(2017)\citenamefont
  {Timounay}, \citenamefont {Pitois},\ and\ \citenamefont
  {Rouyer}}]{Timounay2017}%
  \BibitemOpen
  \bibfield  {author} {\bibinfo {author} {\bibfnamefont {Y.}~\bibnamefont
  {Timounay}}, \bibinfo {author} {\bibfnamefont {O.}~\bibnamefont {Pitois}}, \
  and\ \bibinfo {author} {\bibfnamefont {F.}~\bibnamefont {Rouyer}},\
  }\href@noop {} {\bibfield  {journal} {\bibinfo  {journal} {Physical Review
  Letters}\ }\textbf {\bibinfo {volume} {118}},\ \bibinfo {pages} {228001}
  (\bibinfo {year} {2017})}\BibitemShut {NoStop}%
\bibitem [{\citenamefont {Pitois}\ and\ \citenamefont
  {Rouyer}(2019)}]{Pitois2019}%
  \BibitemOpen
  \bibfield  {author} {\bibinfo {author} {\bibfnamefont {O.}~\bibnamefont
  {Pitois}}\ and\ \bibinfo {author} {\bibfnamefont {F.}~\bibnamefont
  {Rouyer}},\ }\href {\doibase https://doi.org/10.1016/j.cocis.2019.05.004}
  {\bibfield  {journal} {\bibinfo  {journal} {Current Opinion in Colloid \&
  Interface Science}\ }\textbf {\bibinfo {volume} {43}},\ \bibinfo {pages}
  {125} (\bibinfo {year} {2019})}\BibitemShut {NoStop}%
\bibitem [{\citenamefont {Vella}\ and\ \citenamefont
  {Mahadevan}(2005)}]{Vella2005}%
  \BibitemOpen
  \bibfield  {author} {\bibinfo {author} {\bibfnamefont {D.}~\bibnamefont
  {Vella}}\ and\ \bibinfo {author} {\bibfnamefont {L.}~\bibnamefont
  {Mahadevan}},\ }\href@noop {} {\bibfield  {journal} {\bibinfo  {journal}
  {American journal of physics}\ }\textbf {\bibinfo {volume} {73}},\ \bibinfo
  {pages} {817} (\bibinfo {year} {2005})}\BibitemShut {NoStop}%
\bibitem [{\citenamefont {Cicuta}\ and\ \citenamefont
  {Vella}(2009)}]{Cicuta2009}%
  \BibitemOpen
  \bibfield  {author} {\bibinfo {author} {\bibfnamefont {P.}~\bibnamefont
  {Cicuta}}\ and\ \bibinfo {author} {\bibfnamefont {D.}~\bibnamefont {Vella}},\
  }\href@noop {} {\bibfield  {journal} {\bibinfo  {journal} {Physical review
  letters}\ }\textbf {\bibinfo {volume} {102}},\ \bibinfo {pages} {138302}
  (\bibinfo {year} {2009})}\BibitemShut {NoStop}%
\bibitem [{\citenamefont {Proti{\`e}re}\ \emph {et~al.}(2017)\citenamefont
  {Proti{\`e}re}, \citenamefont {Josserand}, \citenamefont {Aristoff},
  \citenamefont {Stone},\ and\ \citenamefont {Abkarian}}]{Protiere2017}%
  \BibitemOpen
  \bibfield  {author} {\bibinfo {author} {\bibfnamefont {S.}~\bibnamefont
  {Proti{\`e}re}}, \bibinfo {author} {\bibfnamefont {C.}~\bibnamefont
  {Josserand}}, \bibinfo {author} {\bibfnamefont {J.~M.}\ \bibnamefont
  {Aristoff}}, \bibinfo {author} {\bibfnamefont {H.~A.}\ \bibnamefont {Stone}},
  \ and\ \bibinfo {author} {\bibfnamefont {M.}~\bibnamefont {Abkarian}},\
  }\href@noop {} {\bibfield  {journal} {\bibinfo  {journal} {Physical review
  letters}\ }\textbf {\bibinfo {volume} {118}},\ \bibinfo {pages} {108001}
  (\bibinfo {year} {2017})}\BibitemShut {NoStop}%
\bibitem [{\citenamefont {Lagarde}\ and\ \citenamefont
  {Proti\`ere}(2020)}]{Lagarde2020}%
  \BibitemOpen
  \bibfield  {author} {\bibinfo {author} {\bibfnamefont {A.}~\bibnamefont
  {Lagarde}}\ and\ \bibinfo {author} {\bibfnamefont {S.}~\bibnamefont
  {Proti\`ere}},\ }\href {\doibase 10.1103/PhysRevFluids.5.044003} {\bibfield
  {journal} {\bibinfo  {journal} {Phys. Rev. Fluids}\ }\textbf {\bibinfo
  {volume} {5}},\ \bibinfo {pages} {044003} (\bibinfo {year}
  {2020})}\BibitemShut {NoStop}%
\bibitem [{\citenamefont {Petit}\ \emph {et~al.}(2016)\citenamefont {Petit},
  \citenamefont {Biance}, \citenamefont {Lorenceau},\ and\ \citenamefont
  {Planchette}}]{Petit2016}%
  \BibitemOpen
  \bibfield  {author} {\bibinfo {author} {\bibfnamefont {P.}~\bibnamefont
  {Petit}}, \bibinfo {author} {\bibfnamefont {A.-L.}\ \bibnamefont {Biance}},
  \bibinfo {author} {\bibfnamefont {E.}~\bibnamefont {Lorenceau}}, \ and\
  \bibinfo {author} {\bibfnamefont {C.}~\bibnamefont {Planchette}},\
  }\href@noop {} {\bibfield  {journal} {\bibinfo  {journal} {Physical Review
  E}\ }\textbf {\bibinfo {volume} {93}},\ \bibinfo {pages} {042802} (\bibinfo
  {year} {2016})}\BibitemShut {NoStop}%
\bibitem [{\citenamefont {Planchette}\ \emph {et~al.}(2018)\citenamefont
  {Planchette}, \citenamefont {Lorenceau},\ and\ \citenamefont
  {Biance}}]{Planchette2018}%
  \BibitemOpen
  \bibfield  {author} {\bibinfo {author} {\bibfnamefont {C.}~\bibnamefont
  {Planchette}}, \bibinfo {author} {\bibfnamefont {E.}~\bibnamefont
  {Lorenceau}}, \ and\ \bibinfo {author} {\bibfnamefont {A.-L.}\ \bibnamefont
  {Biance}},\ }\href {\doibase 10.1039/C8SM00653A} {\bibfield  {journal}
  {\bibinfo  {journal} {Soft Matter}\ }\textbf {\bibinfo {volume} {14}},\
  \bibinfo {pages} {6419} (\bibinfo {year} {2018})}\BibitemShut {NoStop}%
\bibitem [{\citenamefont {Zang}\ \emph {et~al.}(2010)\citenamefont {Zang},
  \citenamefont {Rio}, \citenamefont {Langevin}, \citenamefont {Wei},\ and\
  \citenamefont {Binks}}]{Zang2010}%
  \BibitemOpen
  \bibfield  {author} {\bibinfo {author} {\bibfnamefont {D.}~\bibnamefont
  {Zang}}, \bibinfo {author} {\bibfnamefont {E.}~\bibnamefont {Rio}}, \bibinfo
  {author} {\bibfnamefont {D.}~\bibnamefont {Langevin}}, \bibinfo {author}
  {\bibfnamefont {B.}~\bibnamefont {Wei}}, \ and\ \bibinfo {author}
  {\bibfnamefont {B.}~\bibnamefont {Binks}},\ }\href@noop {} {\bibfield
  {journal} {\bibinfo  {journal} {The European Physical Journal E}\ }\textbf
  {\bibinfo {volume} {31}},\ \bibinfo {pages} {125} (\bibinfo {year}
  {2010})}\BibitemShut {NoStop}%
\bibitem [{\citenamefont {Jambon-Puillet}\ \emph {et~al.}(2017)\citenamefont
  {Jambon-Puillet}, \citenamefont {Josserand},\ and\ \citenamefont
  {Protiere}}]{Jambon2017}%
  \BibitemOpen
  \bibfield  {author} {\bibinfo {author} {\bibfnamefont {E.}~\bibnamefont
  {Jambon-Puillet}}, \bibinfo {author} {\bibfnamefont {C.}~\bibnamefont
  {Josserand}}, \ and\ \bibinfo {author} {\bibfnamefont {S.}~\bibnamefont
  {Protiere}},\ }\href@noop {} {\bibfield  {journal} {\bibinfo  {journal}
  {Physical Review Materials}\ }\textbf {\bibinfo {volume} {1}},\ \bibinfo
  {pages} {042601} (\bibinfo {year} {2017})}\BibitemShut {NoStop}%
\bibitem [{\citenamefont {Saavedra}\ \emph {et~al.}(2018)\citenamefont
  {Saavedra}, \citenamefont {Elettro},\ and\ \citenamefont
  {Melo}}]{Saavedra2018}%
  \BibitemOpen
  \bibfield  {author} {\bibinfo {author} {\bibfnamefont {O.}~\bibnamefont
  {Saavedra}}, \bibinfo {author} {\bibfnamefont {H.}~\bibnamefont {Elettro}}, \
  and\ \bibinfo {author} {\bibfnamefont {F.}~\bibnamefont {Melo}},\ }\href@noop
  {} {\bibfield  {journal} {\bibinfo  {journal} {Physical Review Materials}\
  }\textbf {\bibinfo {volume} {2}},\ \bibinfo {pages} {043603} (\bibinfo {year}
  {2018})}\BibitemShut {NoStop}%
\bibitem [{\citenamefont {He}\ \emph {et~al.}(2020)\citenamefont {He},
  \citenamefont {Sun},\ and\ \citenamefont {Dinsmore}}]{He2020}%
  \BibitemOpen
  \bibfield  {author} {\bibinfo {author} {\bibfnamefont {W.}~\bibnamefont
  {He}}, \bibinfo {author} {\bibfnamefont {Y.}~\bibnamefont {Sun}}, \ and\
  \bibinfo {author} {\bibfnamefont {A.~D.}\ \bibnamefont {Dinsmore}},\ }\href
  {\doibase 10.1039/C9SM01251F} {\bibfield  {journal} {\bibinfo  {journal}
  {Soft Matter}\ }\textbf {\bibinfo {volume} {16}},\ \bibinfo {pages} {2497}
  (\bibinfo {year} {2020})}\BibitemShut {NoStop}%
\bibitem [{\citenamefont {Barman}\ and\ \citenamefont
  {Christopher}(2014)}]{Barman2014}%
  \BibitemOpen
  \bibfield  {author} {\bibinfo {author} {\bibfnamefont {S.}~\bibnamefont
  {Barman}}\ and\ \bibinfo {author} {\bibfnamefont {G.~F.}\ \bibnamefont
  {Christopher}},\ }\href@noop {} {\bibfield  {journal} {\bibinfo  {journal}
  {Langmuir}\ }\textbf {\bibinfo {volume} {30}},\ \bibinfo {pages} {9752}
  (\bibinfo {year} {2014})}\BibitemShut {NoStop}%
\bibitem [{\citenamefont {Barman}\ and\ \citenamefont
  {Christopher}(2016)}]{Barman2016}%
  \BibitemOpen
  \bibfield  {author} {\bibinfo {author} {\bibfnamefont {S.}~\bibnamefont
  {Barman}}\ and\ \bibinfo {author} {\bibfnamefont {G.~F.}\ \bibnamefont
  {Christopher}},\ }\href@noop {} {\bibfield  {journal} {\bibinfo  {journal}
  {Journal of Rheology}\ }\textbf {\bibinfo {volume} {60}},\ \bibinfo {pages}
  {35} (\bibinfo {year} {2016})}\BibitemShut {NoStop}%
\bibitem [{\citenamefont {gdrmidi@ polytech. univ-mrs. fr http://www. lmgc.
  univ-montp2. fr/MIDI/}(2004)}]{GDR2004}%
  \BibitemOpen
  \bibfield  {author} {\bibinfo {author} {\bibfnamefont {G.~M.}\ \bibnamefont
  {gdrmidi@ polytech. univ-mrs. fr http://www. lmgc. univ-montp2. fr/MIDI/}},\
  }\href@noop {} {\bibfield  {journal} {\bibinfo  {journal} {The European
  Physical Journal E}\ }\textbf {\bibinfo {volume} {14}},\ \bibinfo {pages}
  {341} (\bibinfo {year} {2004})}\BibitemShut {NoStop}%
\bibitem [{\citenamefont {Cassar}\ \emph {et~al.}(2005)\citenamefont {Cassar},
  \citenamefont {Nicolas},\ and\ \citenamefont {Pouliquen}}]{Cassar2005}%
  \BibitemOpen
  \bibfield  {author} {\bibinfo {author} {\bibfnamefont {C.}~\bibnamefont
  {Cassar}}, \bibinfo {author} {\bibfnamefont {M.}~\bibnamefont {Nicolas}}, \
  and\ \bibinfo {author} {\bibfnamefont {O.}~\bibnamefont {Pouliquen}},\
  }\href@noop {} {\bibfield  {journal} {\bibinfo  {journal} {Physics of
  fluids}\ }\textbf {\bibinfo {volume} {17}},\ \bibinfo {pages} {103301}
  (\bibinfo {year} {2005})}\BibitemShut {NoStop}%
\bibitem [{\citenamefont {Jop}\ \emph {et~al.}(2006)\citenamefont {Jop},
  \citenamefont {Forterre},\ and\ \citenamefont {Pouliquen}}]{Jop2006}%
  \BibitemOpen
  \bibfield  {author} {\bibinfo {author} {\bibfnamefont {P.}~\bibnamefont
  {Jop}}, \bibinfo {author} {\bibfnamefont {Y.}~\bibnamefont {Forterre}}, \
  and\ \bibinfo {author} {\bibfnamefont {O.}~\bibnamefont {Pouliquen}},\
  }\href@noop {} {\bibfield  {journal} {\bibinfo  {journal} {Nature}\ }\textbf
  {\bibinfo {volume} {441}},\ \bibinfo {pages} {727} (\bibinfo {year}
  {2006})}\BibitemShut {NoStop}%
\bibitem [{\citenamefont {Boyer}\ \emph {et~al.}(2011)\citenamefont {Boyer},
  \citenamefont {Guazzelli},\ and\ \citenamefont {Pouliquen}}]{Boyer2011}%
  \BibitemOpen
  \bibfield  {author} {\bibinfo {author} {\bibfnamefont {F.}~\bibnamefont
  {Boyer}}, \bibinfo {author} {\bibfnamefont {{\'E}.}~\bibnamefont
  {Guazzelli}}, \ and\ \bibinfo {author} {\bibfnamefont {O.}~\bibnamefont
  {Pouliquen}},\ }\href@noop {} {\bibfield  {journal} {\bibinfo  {journal}
  {Physical review letters}\ }\textbf {\bibinfo {volume} {107}},\ \bibinfo
  {pages} {188301} (\bibinfo {year} {2011})}\BibitemShut {NoStop}%
\bibitem [{\citenamefont {Bocquet}\ \emph {et~al.}(2009)\citenamefont
  {Bocquet}, \citenamefont {Colin},\ and\ \citenamefont
  {Ajdari}}]{Bocquet2009}%
  \BibitemOpen
  \bibfield  {author} {\bibinfo {author} {\bibfnamefont {L.}~\bibnamefont
  {Bocquet}}, \bibinfo {author} {\bibfnamefont {A.}~\bibnamefont {Colin}}, \
  and\ \bibinfo {author} {\bibfnamefont {A.}~\bibnamefont {Ajdari}},\
  }\href@noop {} {\bibfield  {journal} {\bibinfo  {journal} {Physical review
  letters}\ }\textbf {\bibinfo {volume} {103}},\ \bibinfo {pages} {036001}
  (\bibinfo {year} {2009})}\BibitemShut {NoStop}%
\bibitem [{\citenamefont {Reddy}\ \emph {et~al.}(2011)\citenamefont {Reddy},
  \citenamefont {Forterre},\ and\ \citenamefont {Pouliquen}}]{Reddy2011}%
  \BibitemOpen
  \bibfield  {author} {\bibinfo {author} {\bibfnamefont {K.}~\bibnamefont
  {Reddy}}, \bibinfo {author} {\bibfnamefont {Y.}~\bibnamefont {Forterre}}, \
  and\ \bibinfo {author} {\bibfnamefont {O.}~\bibnamefont {Pouliquen}},\
  }\href@noop {} {\bibfield  {journal} {\bibinfo  {journal} {Physical Review
  Letters}\ }\textbf {\bibinfo {volume} {106}},\ \bibinfo {pages} {108301}
  (\bibinfo {year} {2011})}\BibitemShut {NoStop}%
\bibitem [{\citenamefont {Bouzid}\ \emph {et~al.}(2013)\citenamefont {Bouzid},
  \citenamefont {Trulsson}, \citenamefont {Claudin}, \citenamefont
  {Cl\'ement},\ and\ \citenamefont {Andr\'eotti}}]{Bouzid2013}%
  \BibitemOpen
  \bibfield  {author} {\bibinfo {author} {\bibfnamefont {M.}~\bibnamefont
  {Bouzid}}, \bibinfo {author} {\bibfnamefont {M.}~\bibnamefont {Trulsson}},
  \bibinfo {author} {\bibfnamefont {P.}~\bibnamefont {Claudin}}, \bibinfo
  {author} {\bibfnamefont {E.}~\bibnamefont {Cl\'ement}}, \ and\ \bibinfo
  {author} {\bibfnamefont {B.}~\bibnamefont {Andr\'eotti}},\ }\href@noop {}
  {\bibfield  {journal} {\bibinfo  {journal} {Phys. Rev. Lett.}\ }\textbf
  {\bibinfo {volume} {111}},\ \bibinfo {pages} {238301} (\bibinfo {year}
  {2013})}\BibitemShut {NoStop}%
\bibitem [{\citenamefont {Bouzid}\ \emph {et~al.}(2015)\citenamefont {Bouzid},
  \citenamefont {Izzet}, \citenamefont {Trulsson}, \citenamefont {Cl{\'e}ment},
  \citenamefont {Claudin},\ and\ \citenamefont {Andreotti}}]{Bouzid2015}%
  \BibitemOpen
  \bibfield  {author} {\bibinfo {author} {\bibfnamefont {M.}~\bibnamefont
  {Bouzid}}, \bibinfo {author} {\bibfnamefont {A.}~\bibnamefont {Izzet}},
  \bibinfo {author} {\bibfnamefont {M.}~\bibnamefont {Trulsson}}, \bibinfo
  {author} {\bibfnamefont {E.}~\bibnamefont {Cl{\'e}ment}}, \bibinfo {author}
  {\bibfnamefont {P.}~\bibnamefont {Claudin}}, \ and\ \bibinfo {author}
  {\bibfnamefont {B.}~\bibnamefont {Andreotti}},\ }\href {\doibase
  10.1140/epje/i2015-15125-1} {\bibfield  {journal} {\bibinfo  {journal} {Eur.
  Phys. J. E}\ }\textbf {\bibinfo {volume} {38}},\ \bibinfo {pages} {125}
  (\bibinfo {year} {2015})}\BibitemShut {NoStop}%
\bibitem [{\citenamefont {Henann}\ and\ \citenamefont
  {Kamrin}(2014)}]{Henann2014}%
  \BibitemOpen
  \bibfield  {author} {\bibinfo {author} {\bibfnamefont {D.~L.}\ \bibnamefont
  {Henann}}\ and\ \bibinfo {author} {\bibfnamefont {K.}~\bibnamefont
  {Kamrin}},\ }\href@noop {} {\bibfield  {journal} {\bibinfo  {journal} {Phys.
  Rev. Lett.}\ }\textbf {\bibinfo {volume} {113}},\ \bibinfo {pages} {178001}
  (\bibinfo {year} {2014})}\BibitemShut {NoStop}%
\bibitem [{\citenamefont {Thomas}\ \emph {et~al.}(2019)\citenamefont {Thomas},
  \citenamefont {Tang}, \citenamefont {Daniels},\ and\ \citenamefont
  {Vriend}}]{Thomas2019}%
  \BibitemOpen
  \bibfield  {author} {\bibinfo {author} {\bibfnamefont {A.}~\bibnamefont
  {Thomas}}, \bibinfo {author} {\bibfnamefont {Z.}~\bibnamefont {Tang}},
  \bibinfo {author} {\bibfnamefont {K.~E.}\ \bibnamefont {Daniels}}, \ and\
  \bibinfo {author} {\bibfnamefont {N.}~\bibnamefont {Vriend}},\ }\href@noop {}
  {\bibfield  {journal} {\bibinfo  {journal} {Soft Matter}\ }\textbf {\bibinfo
  {volume} {15}},\ \bibinfo {pages} {8532} (\bibinfo {year}
  {2019})}\BibitemShut {NoStop}%
\bibitem [{\citenamefont {Gaume}\ \emph {et~al.}(2020)\citenamefont {Gaume},
  \citenamefont {Chambon},\ and\ \citenamefont {Naaim}}]{Gaume2020}%
  \BibitemOpen
  \bibfield  {author} {\bibinfo {author} {\bibfnamefont {J.}~\bibnamefont
  {Gaume}}, \bibinfo {author} {\bibfnamefont {G.}~\bibnamefont {Chambon}}, \
  and\ \bibinfo {author} {\bibfnamefont {M.}~\bibnamefont {Naaim}},\
  }\href@noop {} {\bibfield  {journal} {\bibinfo  {journal} {Physical Review
  Letters}\ }\textbf {\bibinfo {volume} {125}},\ \bibinfo {pages} {188001}
  (\bibinfo {year} {2020})}\BibitemShut {NoStop}%
\bibitem [{\citenamefont {Lois}\ \emph {et~al.}(2008)\citenamefont {Lois},
  \citenamefont {Blawzdziewicz},\ and\ \citenamefont {O'Hern}}]{Lois2008}%
  \BibitemOpen
  \bibfield  {author} {\bibinfo {author} {\bibfnamefont {G.}~\bibnamefont
  {Lois}}, \bibinfo {author} {\bibfnamefont {J.}~\bibnamefont {Blawzdziewicz}},
  \ and\ \bibinfo {author} {\bibfnamefont {C.~S.}\ \bibnamefont {O'Hern}},\
  }\href {\doibase 10.1103/PhysRevLett.100.028001} {\bibfield  {journal}
  {\bibinfo  {journal} {Phys. Rev. Lett.}\ }\textbf {\bibinfo {volume} {100}},\
  \bibinfo {pages} {028001} (\bibinfo {year} {2008})}\BibitemShut {NoStop}%
\bibitem [{\citenamefont {Losert}\ \emph {et~al.}(2000)\citenamefont {Losert},
  \citenamefont {Bocquet}, \citenamefont {Lubensky},\ and\ \citenamefont
  {Gollub}}]{Losert2000}%
  \BibitemOpen
  \bibfield  {author} {\bibinfo {author} {\bibfnamefont {W.}~\bibnamefont
  {Losert}}, \bibinfo {author} {\bibfnamefont {L.}~\bibnamefont {Bocquet}},
  \bibinfo {author} {\bibfnamefont {T.~C.}\ \bibnamefont {Lubensky}}, \ and\
  \bibinfo {author} {\bibfnamefont {J.~P.}\ \bibnamefont {Gollub}},\ }\href
  {\doibase 10.1103/PhysRevLett.85.1428} {\bibfield  {journal} {\bibinfo
  {journal} {Phys. Rev. Lett.}\ }\textbf {\bibinfo {volume} {85}},\ \bibinfo
  {pages} {1428} (\bibinfo {year} {2000})}\BibitemShut {NoStop}%
\bibitem [{\citenamefont {Krieger}(1972)}]{Krieger1972}%
  \BibitemOpen
  \bibfield  {author} {\bibinfo {author} {\bibfnamefont {I.}~\bibnamefont
  {Krieger}},\ }\href@noop {} {\bibfield  {journal} {\bibinfo  {journal} {Adv.
  Colloids Interface Sci.}\ } (\bibinfo {year} {1972})}\BibitemShut {NoStop}%
\bibitem [{\citenamefont {Brady}(1983)}]{Brady1983}%
  \BibitemOpen
  \bibfield  {author} {\bibinfo {author} {\bibfnamefont {J.~F.}\ \bibnamefont
  {Brady}},\ }\href {\doibase DOI: 10.1016/0301-9322(83)90064-2} {\bibfield
  {journal} {\bibinfo  {journal} {International Journal of Multiphase Flow}\
  }\textbf {\bibinfo {volume} {10}},\ \bibinfo {pages} {113 } (\bibinfo {year}
  {1983})}\BibitemShut {NoStop}%
\bibitem [{\citenamefont {Bingham}(1922)}]{Bingham1922}%
  \BibitemOpen
  \bibfield  {author} {\bibinfo {author} {\bibfnamefont {E.~C.}\ \bibnamefont
  {Bingham}},\ }\href@noop {} {\emph {\bibinfo {title} {Fluidity and
  plasticity}}},\ Vol.~\bibinfo {volume} {2}\ (\bibinfo  {publisher}
  {McGraw-Hill},\ \bibinfo {year} {1922})\BibitemShut {NoStop}%
\bibitem [{\citenamefont {Reynaert}\ \emph {et~al.}(2007)\citenamefont
  {Reynaert}, \citenamefont {Moldenaers},\ and\ \citenamefont
  {Vermant}}]{Reynaert2007}%
  \BibitemOpen
  \bibfield  {author} {\bibinfo {author} {\bibfnamefont {S.}~\bibnamefont
  {Reynaert}}, \bibinfo {author} {\bibfnamefont {P.}~\bibnamefont
  {Moldenaers}}, \ and\ \bibinfo {author} {\bibfnamefont {J.}~\bibnamefont
  {Vermant}},\ }\href@noop {} {\bibfield  {journal} {\bibinfo  {journal}
  {Physical Chemistry Chemical Physics}\ }\textbf {\bibinfo {volume} {9}},\
  \bibinfo {pages} {6463} (\bibinfo {year} {2007})}\BibitemShut {NoStop}%
\bibitem [{\citenamefont {Bocquet}\ \emph {et~al.}(2001)\citenamefont
  {Bocquet}, \citenamefont {Losert}, \citenamefont {Schalk}, \citenamefont
  {Lubensky},\ and\ \citenamefont {Gollub}}]{Bocquet2001}%
  \BibitemOpen
  \bibfield  {author} {\bibinfo {author} {\bibfnamefont {L.}~\bibnamefont
  {Bocquet}}, \bibinfo {author} {\bibfnamefont {W.}~\bibnamefont {Losert}},
  \bibinfo {author} {\bibfnamefont {D.}~\bibnamefont {Schalk}}, \bibinfo
  {author} {\bibfnamefont {T.~C.}\ \bibnamefont {Lubensky}}, \ and\ \bibinfo
  {author} {\bibfnamefont {J.~P.}\ \bibnamefont {Gollub}},\ }\href {\doibase
  10.1103/PhysRevE.65.011307} {\bibfield  {journal} {\bibinfo  {journal} {Phys.
  Rev. E}\ }\textbf {\bibinfo {volume} {65}},\ \bibinfo {pages} {011307}
  (\bibinfo {year} {2001})}\BibitemShut {NoStop}%
\bibitem [{\citenamefont {Seguin}\ \emph {et~al.}(2011)\citenamefont {Seguin},
  \citenamefont {Bertho}, \citenamefont {Gondret},\ and\ \citenamefont
  {Crassous}}]{Seguin2011}%
  \BibitemOpen
  \bibfield  {author} {\bibinfo {author} {\bibfnamefont {A.}~\bibnamefont
  {Seguin}}, \bibinfo {author} {\bibfnamefont {Y.}~\bibnamefont {Bertho}},
  \bibinfo {author} {\bibfnamefont {P.}~\bibnamefont {Gondret}}, \ and\
  \bibinfo {author} {\bibfnamefont {J.}~\bibnamefont {Crassous}},\ }\href
  {\doibase 10.1103/PhysRevLett.107.048001} {\bibfield  {journal} {\bibinfo
  {journal} {Phys. Rev. Lett.}\ }\textbf {\bibinfo {volume} {107}},\ \bibinfo
  {pages} {048001} (\bibinfo {year} {2011})}\BibitemShut {NoStop}%
\bibitem [{\citenamefont {Guyon}\ \emph {et~al.}(2001)\citenamefont {Guyon},
  \citenamefont {Hulin}, \citenamefont {Petit}, \citenamefont {Mitescu} \emph
  {et~al.}}]{Guyon2001}%
  \BibitemOpen
  \bibfield  {author} {\bibinfo {author} {\bibfnamefont {E.}~\bibnamefont
  {Guyon}}, \bibinfo {author} {\bibfnamefont {J.-P.}\ \bibnamefont {Hulin}},
  \bibinfo {author} {\bibfnamefont {L.}~\bibnamefont {Petit}}, \bibinfo
  {author} {\bibfnamefont {C.~D.}\ \bibnamefont {Mitescu}},  \emph {et~al.},\
  }\href@noop {} {\emph {\bibinfo {title} {Physical hydrodynamics}}}\ (\bibinfo
   {publisher} {Oxford university press},\ \bibinfo {year} {2001})\BibitemShut
  {NoStop}%
\bibitem [{\citenamefont {Blanc}\ \emph {et~al.}(2011)\citenamefont {Blanc},
  \citenamefont {Peters},\ and\ \citenamefont {Lemaire}}]{Blanc2011}%
  \BibitemOpen
  \bibfield  {author} {\bibinfo {author} {\bibfnamefont {F.}~\bibnamefont
  {Blanc}}, \bibinfo {author} {\bibfnamefont {F.}~\bibnamefont {Peters}}, \
  and\ \bibinfo {author} {\bibfnamefont {E.}~\bibnamefont {Lemaire}},\
  }\href@noop {} {\bibfield  {journal} {\bibinfo  {journal} {Applied Rheology}\
  }\textbf {\bibinfo {volume} {21}} (\bibinfo {year} {2011})}\BibitemShut
  {NoStop}%
\bibitem [{\citenamefont {Kamrin}(2019)}]{Kamrin2019}%
  \BibitemOpen
  \bibfield  {author} {\bibinfo {author} {\bibfnamefont {K.}~\bibnamefont
  {Kamrin}},\ }\href@noop {} {\bibfield  {journal} {\bibinfo  {journal}
  {Frontiers in Physics}\ }\textbf {\bibinfo {volume} {7}},\ \bibinfo {pages}
  {116} (\bibinfo {year} {2019})}\BibitemShut {NoStop}%
\bibitem [{\citenamefont {Seguin}\ and\ \citenamefont
  {Gondret}(2017)}]{Seguin2017}%
  \BibitemOpen
  \bibfield  {author} {\bibinfo {author} {\bibfnamefont {A.}~\bibnamefont
  {Seguin}}\ and\ \bibinfo {author} {\bibfnamefont {P.}~\bibnamefont
  {Gondret}},\ }\href@noop {} {\bibfield  {journal} {\bibinfo  {journal}
  {Physical Review E}\ }\textbf {\bibinfo {volume} {96}},\ \bibinfo {pages}
  {032905} (\bibinfo {year} {2017})}\BibitemShut {NoStop}%
\bibitem [{\citenamefont {Tighe}\ \emph {et~al.}(2010)\citenamefont {Tighe},
  \citenamefont {Woldhuis}, \citenamefont {Remmers}, \citenamefont {van
  Saarloos},\ and\ \citenamefont {van Hecke}}]{Tighe2010}%
  \BibitemOpen
  \bibfield  {author} {\bibinfo {author} {\bibfnamefont {B.~P.}\ \bibnamefont
  {Tighe}}, \bibinfo {author} {\bibfnamefont {E.}~\bibnamefont {Woldhuis}},
  \bibinfo {author} {\bibfnamefont {J.~J.}\ \bibnamefont {Remmers}}, \bibinfo
  {author} {\bibfnamefont {W.}~\bibnamefont {van Saarloos}}, \ and\ \bibinfo
  {author} {\bibfnamefont {M.}~\bibnamefont {van Hecke}},\ }\href@noop {}
  {\bibfield  {journal} {\bibinfo  {journal} {Physical review letters}\
  }\textbf {\bibinfo {volume} {105}},\ \bibinfo {pages} {088303} (\bibinfo
  {year} {2010})}\BibitemShut {NoStop}%
\bibitem [{\citenamefont {Chaudhuri}\ \emph {et~al.}(2012)\citenamefont
  {Chaudhuri}, \citenamefont {Berthier},\ and\ \citenamefont
  {Bocquet}}]{Chaudhuri2012}%
  \BibitemOpen
  \bibfield  {author} {\bibinfo {author} {\bibfnamefont {P.}~\bibnamefont
  {Chaudhuri}}, \bibinfo {author} {\bibfnamefont {L.}~\bibnamefont {Berthier}},
  \ and\ \bibinfo {author} {\bibfnamefont {L.}~\bibnamefont {Bocquet}},\ }\href
  {\doibase 10.1103/PhysRevE.85.021503} {\bibfield  {journal} {\bibinfo
  {journal} {Phys. Rev. E}\ }\textbf {\bibinfo {volume} {85}},\ \bibinfo
  {pages} {021503} (\bibinfo {year} {2012})}\BibitemShut {NoStop}%
\end{thebibliography}%

\end{document}